# Gain lifetime characterization through time-resolved stimulated emission in a whispering-gallery mode microresonator

Xiao-Fei Liu[1], Fuchuan Lei[3], Tie-Jun Wang[1], Gui-Lu Long[2,4,*], and Chuan Wang[1,5,*]

[1] *State Key Laboratory of Information Photonics and Optical Communications and School of Science, Beijing University of Posts and Telecommunications, Beijing 100876, China*

[2] *State Key Laboratory of Low-dimensional Quantum Physics and Department of Physics, Tsinghua University, Beijing 100084, China*

[3] *Light-Matter Interactions Unit, Okinawa Institute of Science and Technology Graduate University, Onna, Okinawa 904-0495, Japan*

[4] *Beijing Academy of Quantum Information, Beijing 100085, China*

[5] *College of Information Science and Technology, Beijing Normal University, Beijing 100875, China*

*Corresponding authors: gllong@tsinghua.edu.cn; wangchuan@bupt.edu.cn

The precise measurement of gain lifetime at a specific wavelength holds significant importance for understanding the properties of photonic devices and further improving their performances. Here, we show that the evolution of gains can be well characterized by measuring linewidth changes of an optical mode in a microresonator; this method cannot be achieved using time-resolved photoluminescence (PL) spectroscopy. We use an erbium-doped high-Q whispering-gallery microresonator to show the feasibility of this method. With the increase of time after the pump laser is turned off, the transmission spectrum of a probe signal exhibits transitions from a Lorentz peak to a dip; this indicates a decay of optical gains, and the corresponding lifetime is estimated to be 5.1 ms. Moreover, taper fiber coupling is used to increase the pump and collection efficiency. This method can be extended to other materials and nanostructures.

## I. INTRODUCTION

Optical gain can dramatically modify the response of a system and compensate for the loss of light during communications and information processing. Such gain processes play an important role in various applications [1], including lasers and masers, optical amplifiers, and sensitive detections. The optical gain typically results from the stimulated emission process, which is generated from the recombination of electrons and holes in semiconductors or the coherent amplification through population inversion between different energy levels in gain medium. The lifetime of optical gain, a fundamental parameter and of significant importance for understanding light-matter interaction, determines various properties of a system, for example, the maximum switch speed of input signal in an active optical switch [2]. The stimulated radiation process corresponds to specific energy level transitions. Therefore, the emission light is considered to have same properties as those of the input light. By contrast, the luminescence, which results from the spontaneous radiation of excited states and multi-channel transitions, has a wide broadening in spectrum and contains lights in different wavelengths [3]. The lifetime of luminescence can be accurately measured through the changes of its intensity after the pump source is turned off; also, optical gain lifetime needs a simple, effective, and fast method to be characterized precisely.

The whispering-gallery mode (WGM) microresonator [4-6], which has an ultra-high quality factor higher than $10^8$ and mode volumes at the cubic micrometer scale, has been widely used in many applications [7-13]. Moreover, the WGM microresonator serves as a fruitful platform to study various applications of optical gains. Ultra-low-threshold micro-lasers have been implemented using a silica microsphere, microdisk, and microtoroid [14, 15]. Furthermore, sensitive metrology and detection have been realized in active microresonators [16-20]. WGM microresonators have received strong interest for the realization of cavity quantum electrodynamical systems [7-8] and the classical analog of quantum systems, such as parity-time optics [21-25] and electromagnetically induced transparency [26-30]. To further improve the performance of WGM microresonators, one can decrease their sizes to obtain smaller mode volumes and stronger light-matter interactions; however, this approach can increase bending losses and lower Q values. Another method is to dope rare-earth ions (e.g., $Er^{3+}$, $Yb^{3+}$, or $Nd^{3+}$) [1, 3, 14] or introduce intrinsic Raman gain from the material [17, 18, 31]. For rare-earth ions (erbium ions, for example), transitions between internal energy levels are stable because these processes taken place in 4f shells are shielded from the interaction with other atoms by the external 5s and 5p electrons. This stability makes rare-earth ions a universal candidate for practical applications, e.g., displays or lighting, performance improvements of magnetic materials, and optical communications. Recently, rare-earth-doped WGM microresonators have been successfully applied in signal amplifiers [32], ultra-long storage of photons [33], and on-demand coupling control [34, 35].

To date, there is no related study that has been conducted to identify optical gain lifetime in microresonators. Here, we present a direct and precise measurement of this lifetime at a specific wavelength through the linewidth changes of an optical mode. An $Er^{3+}$-doped material is taken as an example to demonstrate the feasibility of this method. To enhance the light-matter interaction, we dope erbium ions into WGM microresonators. In our study, a pump laser with central wavelength of 1430 nm and input power of 275 µW is coupled into the microtoroid through a tapered fiber to excite erbium ions into intermediate states. Then, this excitation is turned off by optothermal squeezing within 2 µs. Next, another laser in 1550 nm band with a high scanning speed of 4.2 THz/s serves as the probe signal. To eliminate the influence of temperature, the thermal relaxation time is also precisely characterized to be 0.33 ms via optothermal spectroscopy. We use six different probe signals with the wavelength scanned over the range from 1528 nm to 1556 nm and wavelength differences of 5.6 nm, which is approximately one free spectral range (FSR). Gain lifetime is measured to be approximately 5.1 ms. Through continually tuning the microresonator, lifetime of optical gains at different resonant wavelengths can also been well measured. Our results precisely show the evolution of optical gain at specific wavelengths, and the lifetime can be characterized simultaneously by measuring the linewidth changes of a probe mode; this method cannot be achieved by using the standard time-resolved photoluminescence (PL) spectroscopy technique. Moreover, the use of tapered fiber coupling allows for higher pump and collection efficiencies, thereby greatly decreasing the demand for detectors and measurement techniques

## II. THE METHODS AND THE EVOLUTION OF TRANSMISSION SPECTRA

The experimental configuration is depicted in Fig. 1, which shows a silica microtoroidal WGM resonator directly coupled to a tapered fiber. The WGM resonator is an $Er^{3+}$-doped microtoroid, which is fabricated using the sol-gel technique [21, 36]. First, silica layers with a thickness of 2 µm and an ion concentration of $7.4\times10^{24}$ ions/cm$^3$ are fabricated on the surface of a silicon wafer. Subsequently, photolithography, pattern transfer, dry etching, and reflow are applied to form this surface-tension-induced microtoroid. The WGM resonator has intrinsic Q values of $4.2\times10^6$ in the 1430 nm pump mode and $5.3\times10^6$ in the 1550 nm probe mode. The tapered fiber [37, 38], with a waist diameter of approximately 2 μm, is fabricated through the heat and pull process using a single-mode optical fiber operating in the 1550 nm window. The pump laser and probe signal are provided by two tunable laser diodes with linewidths less than 200 kHz. The central wavelengths and scanning speeds of these laser diodes can be precisely controlled by a computer. The light emitted from laser diodes is coupled into the microtoroid through a coupler and tapered fiber. This tapered fiber has a coupling efficiency exceeding 99%, i.e., the normalized transmission can nearly reach zero and more than 99 % of input light can couple into the microtoroid when an optical mode is excited resonantly. Moreover, a three-axis stage with resolution 0.1 μm can continuously tune the coupling strength between this taper fiber and the microtoroid. The light coupled out is divided into two parts. One component (constituting 10 %) part is connected with an optical spectrum analyzer (OSA), and the other component (constituting the remaining 90 % part) is further divided into 1550 nm and 1430 nm bands after passing through a wavelength division multiplexer (WDM). Finally, the light from these two bands are collected by two photon detectors (PDs) and monitored by an oscilloscope. During the entire process, two variable optical attenuators (VOAs) and polarization controllers (PCs) are used to control the input power and polarization, respectively.

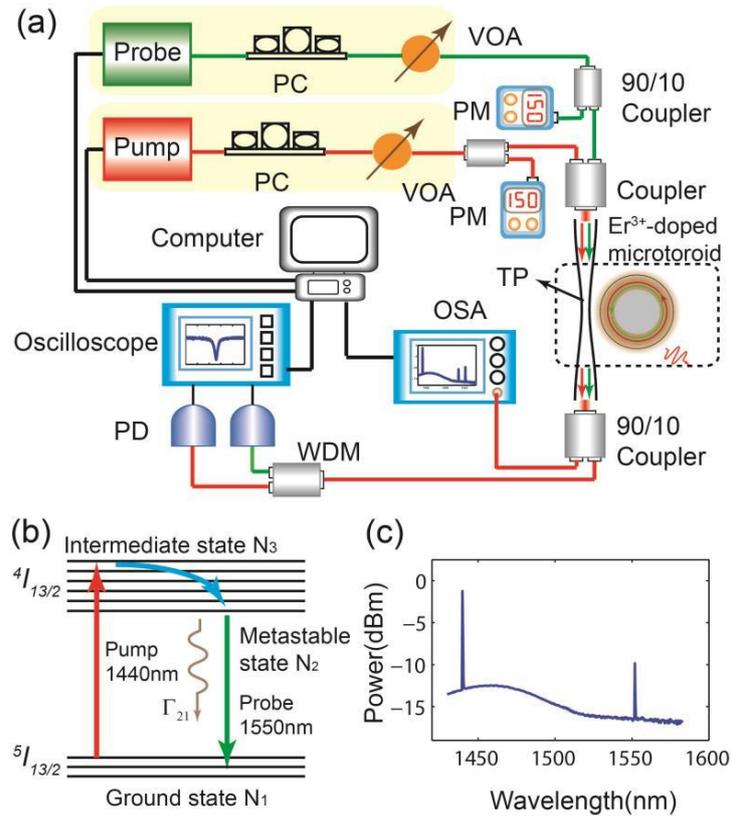

Fig. 1. (a) Experimental configuration for measuring optical gain lifetime. VOA, variable optical attenuator; WDM, wavelength division multiplexer; PD, photodetector; OSA, optical spectrum analyzer; PM, power meter; TP, tapered fiber; PC, polarization controller. (b) The energy levels of erbium ions. (c) Emission spectra of $Er^{3+}$-doped WGM microcavities pumped in the 1430 nm band.

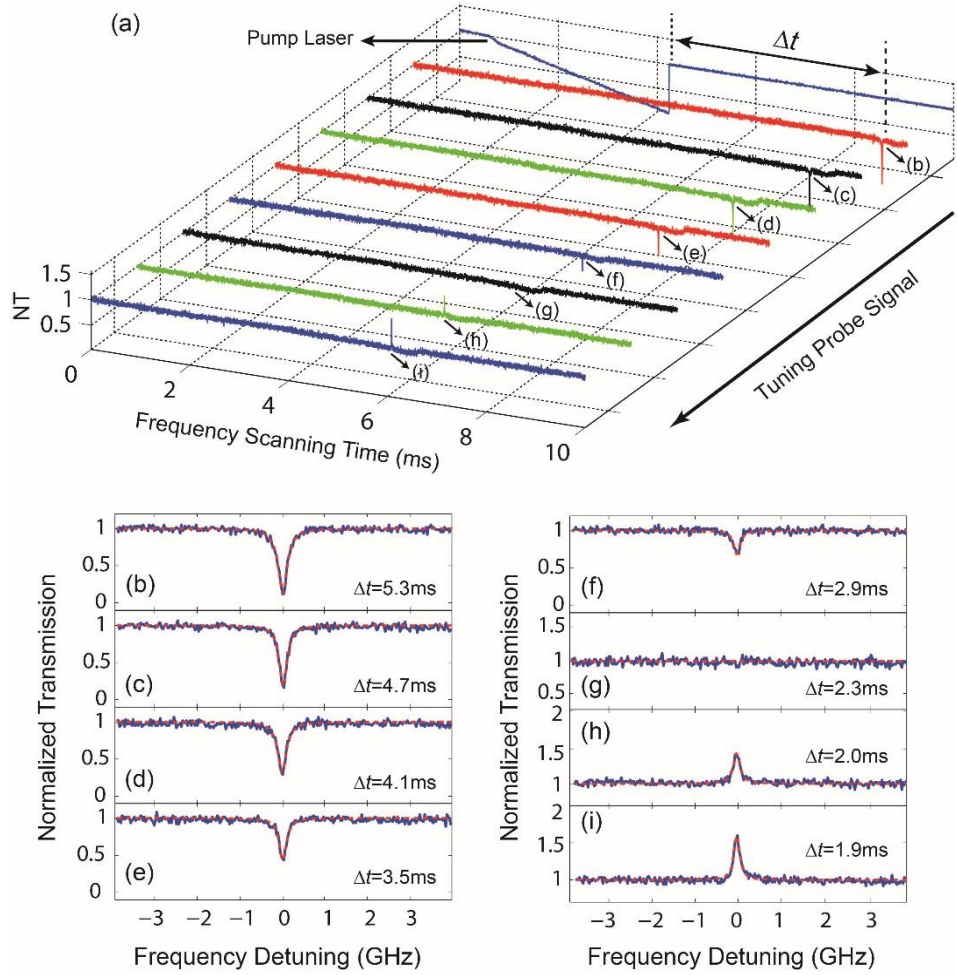

Fig. 2. (a) Evolution of the transmission spectra when measuring the optical gain lifetime. When we decrease the time $\Delta t$ after the pump laser is turned off, this probe signal become increasingly amplified together with more gains. (b)-(i) show the details and the fitting curves (red dashed lines) in (a). Note that the horizontal coordinate of (a), i.e., frequency scanning time, has been transformed into frequency detuning in (b)-(i) according to the scanning speed of this probe signal. The pump power here is 275 μW, and the probe signal power is 150 nm. NT denotes normalized transmission.

Figure 2 demonstrates the process to measure the lifetime of optical gain. Here, the pump laser in 1430 nm band is linearly modulated at a lower frequency scanning speed 1.2 THz/s, and the frequency scanning speed of the probe signal in 1550 nm band is 4.2 THz/s. The transmission spectrum of this pump laser exhibits triangular broadening due to the optothermal effect, and the probe signal has a standard Lorentz lineshape. Even after being turned off, the pump laser can still provide optical gain for this probe signal after the delay time $\Delta t$. Here, we turn off the pump laser within 2 μs through thermal mode squeezing. This mode squeezing is induced by the optothermal effect, where the central wavelength of this pump mode has an opposite shift direction with respect to the input light [29]. As shown in Fig. 2(a), we continuously tune this probe signal by decreasing $\Delta t$ while the pump laser is kept unchanged. The transmission spectra of this probe signal evolve from Lorentz dip to Lorentz peak. Note that the measurement in this figure is in the over coupling regime, i.e., the coupling strength between the tapered fiber and microtoroid is greater than the intrinsic dissipation rate of the cavity mode. Figures 2(b) to 2(i) exhibit the changes of these transmission spectra in details. This process can be divided into three regions: the Lorentz dip region, the transparency region, and the Lorentz peak region. Figures 2(b) to 2(f) correspond to the Lorentz dip region. In this region, the linewidth of this probe signal becomes increasingly small with the decrease of $\Delta t$, and the minimum transmission becomes increasingly high. At the time $\Delta t = 2.3$ ms, the transmission becomes totally transparent, indicating that the existence of this microtoroid does not cause any loss of the light; that is to say, optical gain provided by erbium ions can completely compensate the intrinsic loss. The Lorentz peak region is shown in Figs. 2(h) and 2(i), in which this probe signal is amplified and the transmission is above the normalized line. In this Lorentz peak region, the linewidth of this resonance would continue to decrease, and the peak value becomes increasingly high with the decrease of $\Delta t$. The evolution of this probe signal can be described by the differential equation based on coupled-mode theory [39, 40]:

$$\frac{da_s}{dt} = \left[i\Delta\omega_s(t) - \frac{\kappa_s^{ext} + \kappa_s^0 - g_s}{2}\right]a_s - \sqrt{\kappa_s^{ext}}a_s^{in}, \quad (1)$$

where $a_s$ describes the amplitude of this probe signal inside the microtoroid, $\left|a_s^{in}\right|^2$ denotes the input power, $\kappa_s^0$ is the intrinsic energy decay rate, and $\kappa_s^{ext}$ represents the coupling strength between microtoroid and taper fiber. $g_s$ is the optical gain provided by the transition of erbium ions from metastable states into ground states. $\Delta\omega_s(t)$ is the detuning between the input signal and the central frequency of this probe mode. Since this probe signal is linearly modulated at a high scanning speed, $g_s$ can be regarded as a constant when its input power is small enough (Supplementary Fig. S6).

Under the steady-state situation, the normalized transmission can be expressed by the following equation using the standard input-output relationship:

$$T = \left|1 + \frac{\kappa_s^{ext}}{i\Delta\omega_s - (\kappa_s^{ext} + \kappa_s^0 - g_s)/2}\right|^2. \quad (2)$$

Here, $\kappa_s^0 - g_s$ can be rewritten as $\kappa_s^{eff}$, which denotes the effective energy decay rate after considering the optical gain provided by erbium ions. This transmission is totally transparent under the condition that $\kappa_s^0 = g_s$, while it exhibits a Lorentz dip (or peak) when meeting the condition $\kappa_s^0 > g_s$ (or $\kappa_s^0 < g_s$).

Note that when the pump laser is turned off, the temperature of microtoroid follows an exponential decay. This decay may result in mode squeezing and cause errors in the measurement of optical gains. Therefore, the delay time $\Delta t$ should be greater than the thermal relaxation time $1/\gamma_T$ to avoid this optothermal effect. Here, the thermal relaxation time $1/\gamma_T$ of the microtoroid is characterized to be 0.33 ms using the optothermal spectroscopy method (more details are introduced in Supplementary Note 1). During the gain lifetime measurement, the temperature of the microtoroid is always the same as room temperature. Moreover, $\kappa_s^{ext}$ and $\kappa_s^0$ are kept unchanged. Therefore, changes in $g_s$ tend to modify the transmission spectra.

### III. MEASUREMENT OF GAIN LIFETIME

Figure 1(b) shows the energy levels of erbium ions in a standard Λ-type level system. For convenience, the $^5I_{13/2}$ state is rewritten as the ground state, and the $^4I_{13/2}$ state consists of both the intermediate state and metastable state. The pump laser in the 1430 nm band excites erbium ions from the ground state into intermediate state; subsequently, this excitation decays into the metastable state quickly without the emission of photons. In addition, erbium ions in metastable states decay into ground states at a relatively low rate $\Gamma_{21}$; the lifetime $1/\Gamma_{21}$ can be as long as several milliseconds. When a rapidly modulated probe signal (which can be regarded as a pulse in the time domain with a width of 15.1 μs) in the 1550 nm band is coupled into this system, erbium ions in metastable states would jump into ground states and provide optical gain for this probe signal. The gain $g_s$ is determined by the population inversions between these two energy levels [1]:

$$g_s = \frac{c}{n_s}(\sigma_s^e N_2 - \sigma_s^a N_1). \qquad (3)$$

Here, $\sigma_s^e$ and $\sigma_s^a$ are the emission and absorption cross sections of a probe signal, respectively; $c$ is the speed of light; $n_s$ denotes the effective refractive index; $N_2$ and $N_1$ are the population of erbium ions in the metastable state and ground state within a unit volume, respectively. The condition $g_s < 0$ indicates that the probe signal is absorbed by erbium ions and cannot achieve amplification. If and only if the number of effective erbium ions in the metastable state is greater than that in the ground state, i.e., $N_2 > N_1 \sigma_s^a / \sigma_s^e$, can the compensation of loss and signal amplification be achieved. Assuming that erbium ions in metastable states decay into ground states at the spontaneous decay rate $\Gamma_{21}$, i.e., $N_2(t) = N_2(0)e^{-\Gamma_{21}\Delta t}$; when the time $\Delta t$ is infinitely long, then all erbium ions are located in ground states. Moreover, this system satisfies the relationship $N_1 + N_2 = N_{total}$ after the pump laser is turned off. According to Eq. (3), the energy decay rate $\kappa_{Er^{3+}}$ caused by the absorption of erbium ions if all ions are in ground states could be written as $\kappa_{Er^{3+}} = c\sigma_s^a N_{total}/n_s$.

From Eq. (3) and above analyses, the effective energy decay rate $\kappa_s^{eff} = \kappa_s^0 - g_s$, which is obtained from the transmission spectrum, can be rewritten as:

$$\kappa_s^{eff} = \kappa_s - g_{max}e^{-\Gamma_{21}\Delta t}. \qquad (4)$$

In the above expression, $\kappa_s = \kappa_s^0 + \kappa_{Er^{3+}}$, which represents the effective intrinsic losses after considering the absorption of light by erbium ions. $g_{max} = c(\sigma_s^a + \sigma_s^e)N_2(0)/n_s$, and $g_s$ can be rewritten as $g_s = g_{max}e^{-\Gamma_{21}\Delta t} - \kappa_{Er^{3+}}$. When we have obtained the transmission spectra of probe signal at different time $\Delta t$, $\kappa_s^{eff}$ obeys an exponential decay at the rate of $\Gamma_{21}$ through the translation transformation by deleting $\kappa_s$.

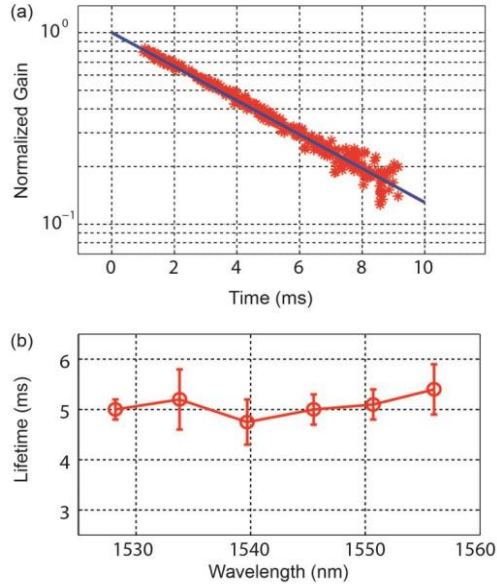

Fig. 3.  (a) Normalized optical gains $(\kappa_s^{eff} - \kappa_s)/g_{max}$ for a probe signal with input power 150 nW. The parameters after fitting are $\kappa_s = 0.189$ GHz, and $g_{max} = 0.299$ GHz. (b) Lifetimes of optical gains at various probe wavelengths.

The experimental results are shown in Fig. 3. As shown in Fig. 3(a), gain lifetime at the resonant wavelength 1551 nm is 5.1 ms when the input power of this probe signal is 150 nW; furthermore, if the input power is increased to 230 nW and 310 nW, nearly the same value is obtained (see Supplementary Fig. S5). Note that $g_s$ can be regarded as a constant only when the input power is small enough; otherwise, the Fano-like resonance shape may appear. Figure 3(b) demonstrates experimental results at various resonant wavelengths, which are all around 5.1 ms. From Eq. (4), the lifetime of optical gain is determined only by the spontaneous decay rate from the $^4I_{13/2}$ state to the $^5I_{13/2}$ state. Meanwhile, the strength of optical gains is also determined by the absorption and emission cross sections $\sigma_s^e$ and $\sigma_s^a$ at a specific wavelength. Note that the lifetime of optical gains $1/\Gamma_{21}$ within this ultrahigh Q microresonator is nearly the same as that in free space, i.e., the Purcell factor [41], which describes the enhancement of spontaneous emission inside a resonator, is relatively small. From Fermi's golden rule, the decay rate $\Gamma_{21}$ is determined by the local density of states (LDOS) of ground states. The emission spectrum of erbium ions is much larger than the linewidth of an optical mode in microtoroid. As a result, when the transition from the $^4I_{13/2}$ state to the $^5I_{13/2}$ state can coincide with one optical mode, the LDOS of corresponding ground state is greatly enhanced. However, for other transitions that cannot match with optical modes are forbidden, because the corresponding LDOS are zero. Therefore, the average effect of erbium ions is nearly the same as that in free space.

The population of erbium ions follows a Boltzmann distribution when the system is not pumped, i.e., $N_2/N_1 = e^{-\hbar\omega_s/k_BT}$, and $N_3/N_1 = e^{-\hbar\omega_p/k_BT}$. Since the density of erbium ions is fixed, the relationship $N_1 + N_2 + N_3 = N_{total}$ holds. According to laser rate equations, the evolution is controlled by the following formulas when we consider only the effect of pump laser:

$$\frac{dN_2}{dt} = \Gamma_{32}N_3 - \Gamma_{21}N_2, \tag{5}$$

$$\frac{dN_3}{dt} = -\Gamma_{32}N_3 + \phi_p(\sigma_p^e N_1 - \sigma_p^a N_3). \tag{6}$$

where $\Gamma_{32}$ is the transition rate of erbium ions from the intermediate state into the metastable state; this rate is generally larger than $\Gamma_{21}$ by three orders of magnitude. $\phi_p$ is the photon flux, i.e., the number of photons per unit area within a unit time. For WGM microcavities, the photon flux can be expressed as $c|a_p|^2/(\hbar\omega_p V_p n_p)$, in which $V_p$ and $n_p$ are the mode volume and refractive index of the pump mode, respectively. For simplicity, we can obtain the behavior of erbium ions quantitatively in steady-state situations by setting the above equations to zero. Furthermore, under the condition $\Gamma_{32} \gg \Gamma_{21}$, it is straightforward to determine that $N_3/N_2 = \Gamma_{21}/\Gamma_{32}$ and $N_2 \approx N_{total}\sigma_p\phi_p/(\Gamma_{21} + \sigma_p\phi_p)$ (Supplementary Note 2). With increasing pump power, more erbium ions are in the metastable state, and less are in the ground state. Since the Q factor of the pump mode is higher than $10^6$, a weak pump can lead to the saturation of the excited states, i.e., $N_2 \approx N_{total}$. This is the origin of many nonlinear behaviors in WGM microcavities.

Figure 4 depicts the behavior of microtoroid and erbium ions using numerical calculations. According to the temperature changes, this figure can be divided into three regions. In Region 1, when little pump laser is coupled into microtoroid, most erbium ions are in their ground states. Moreover, the temperature remains nearly the same as room temperature. In Region 2, more pump light is coupled into the microtoroid with the decrease of detuning $\Delta\omega_p$, and erbium ions start to make transitions from the ground state into the intermediate state. Furthermore, this region can be divided into saturated and unsaturated areas. In the unsaturated area, population inversions increase sharply. In the saturated area, where the pump is strong enough to satisfy the condition $\sigma_p \phi_p \gg \Gamma_{21}$, nearly all the erbium ions are in the metastable state, i.e., $N_2 \approx N_{total}$. In Region 3, when the pump laser is turned off, both the temperature and the number of erbium ions in metastable states exhibit an exponential decay. These two rates obey the relation $\gamma_T/\Gamma_{21} \approx 15.45$. For $N_2 > N_1 \sigma_s^a/\sigma_s^e$, optical gain satisfies the condition $g_s > 0$, as shown in blue. In contrast, erbium ions would absorb the input signal when $N_2 < N_1 \sigma_s^a/\sigma_s^e$ and $g_s < 0$; this condition could cause additional losses of light, as shown in gray.

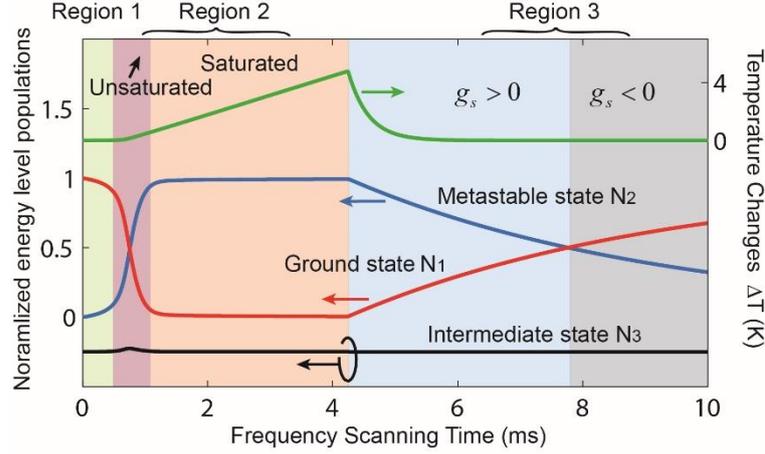

Fig. 4.   Populations of erbium ions in ground states $N_1$ (red line), metastable states $N_2$ (blue line), and intermediate states $N_3$ (black line), and temperature changes $\Delta t$ (green line). Note that the intermediate state population is moved down by 0.25. The parameters are $V_{mode} = 367 \mu m^3$, $\sigma_p^e = \sigma_p^a = 1.5 \times 10^{-22} cm^2$, $1/\Gamma_{32} = 9.2\mu s$, $1/\Gamma_{21} = 5.1$ ms, $N_{total} = 7.3 \times 10^{18}$ ions/cm³, $|a_p^{in}|^2 = 277\mu W$, $\gamma_p = 1.97 \times 10^4$ W/K, $1/\gamma_T = 0.33$ ms, and $\tau_c = 1.17$ ps.

Since erbium ions are doped into the microresonator, the decay rate $\Gamma_{21}$ contains the Purcell effect enhanced spontaneous decay rate $F_p \Gamma_r$, the erbium ion energy exchange induced decay rate $\Gamma_{Er^{3+}}$, and the non-radiative decay rate $\Gamma_{non}$. Strictly speaking, the lifetime of optical gains and that of fluorescence are different. In this simplified model, we assume that the metastable state is isolated; therefore, they share the same value. However, the metastable state is usually a continuum for real erbium ions, and energy relaxations occur between the internal energy levels. The decay of fluorescence originates from the average transitions from the continuum metastable state into the ground state, and the decay of optical gains corresponds only to the specific transitions between the metastable state and the ground state. Therefore, the lifetime of optical gains and the fluorescence are not exactly the same, although their values may be nearly the same. Our work provides a practical and simple method to precisely measure the lifetime of optical gains at specific wavelengths; such measurements cannot be achieved by the previous PL method. Moreover, this method can reflect the value of the metastable state if neglecting its internal relaxations. Furthermore, when the resonant wavelength of this microtoroid is continuously tuned, the lifetime of optical gain at different wavelengths can also be measured. The tapered fiber coupling used here improves this method's pump and collection efficiency, which can greatly decrease the demands for detectors and other measurement techniques.

## IV. CONCLUSION

In this article, we present the direct measurement of gain lifetime at a specific wavelength through the changes of one optical mode; such a measurement can not be achieved by the previous PL method. An erbium-doped WGM microresonator was used to show the efficiency of this method. We used six probe signals with the wavelength differences of 5.6 nm (one FSR). All the optical gain lifetime values were found to be approximately 5.1 ms, which is much longer than the thermal relaxation time of 0.33 ms. The method is universal and can be extended to other materials and structures.

## ACKNOWLEDGEMENTS

The authors gratefully acknowledge the support from the Ministry of Science and Technology of the People's Republic of China (MOST) (grant no. 2016YFA0301304); In part by the Beijing Advanced Innovation Center for Future Chip (ICFC); The National Natural Science Foundation of China (grant nos. 61622103, 61727801, 11774197, 61471050 and 61671083); The Fok Ying-Tong Education Foundation for Young Teachers in the Higher Education Institutions of China (grant no. 151063); And the Fund of the State Key Laboratory of Information Photonics and Optical Communications (Beijing University of Posts and Telecommunications), P. R. China.

----------------------------------------------

[1] P. M. Becker, A. A. Olsson, and J. R. Simpson, "Erbium-doped fiber amplifiers: fundamentals and technology." (Academic press, 1999).

[2] V. S. Ilchenko, and A. B. Matsko, "Optical resonators with whispering-gallery modes-part ii: applications." IEEE J. Sel. Top. Quantum Electron. 12, 15-32 (2006).

[3] T. Reynolds, N. Riesen, A. Meldrum, X. Fan, J. M. M. Hall, T. M. Monro, and A. François, "Fluorescent and lasing whispering gallery mode microresonators for sensing applications." Laser Photon. Rev. 11, 1600265 (2017).

[4] K. J. Vahala, "Optical microcavities." Nature 424, 839-846 (2003).

[5] A. Chiasera, Y. Dumeige, P. Féron, M. Ferrari, Y. Jestin, G. N. Conti, S. Pelli, S. Soria, and G. C. Righini "Spherical whispering-gallery-mode microresonators." Laser Photon. Rev. 4, 457-482 (2010).

[6] D. K. Armani, T. J. Kippenberg, S. M. Spillane, and K. J. Vahala, "Ultra-high-Q toroid microcavity on a chip." Nature 421, 925-928 (2003).

[7] B. Dayan, A. S. Parkins, T. Aoki, E. P. Ostby, K. J. Vahala, and H. J. Kimble, "A Photon Turnstile Dynamically Regulated by One Atom." Science 319, 1062-1065 (2008).

[8] M. Scheucher, A. Hilico, E. Will, J. Volz, and A. Rauschenbeutel, "Quantum optical circulator controlled by a single chirally coupled atom." Science 354, 1577--1580 (2016).

[9] Y. Zhi, X.-C. Yu, Q. Gong, L. Yang, and Y.-F. Xiao, "Single Nanoparticle Detection Using Optical Microcavities." Adv. Mater. 29, 1604920 (2017).

[10] Monifi, F., Zhang, J., Ş. K. Özdemir, Peng, B., Liu, Y.-X. Bo, F., Nori, and L.Yang, "Optomechanically induced stochastic resonance and chaos transfer between optical fields." Nature Photon. 10, 399-405 (2016).

[11] J. Zhu, Ş. K. Özdemir, and L. Yang, "Infrared light detection using a whispering-gallery-mode optical microcavity." Appl. Phys. Lett. 104, 171114 (2014).

[12] T. J. Kippenberg, R. Holzwarth, and S. A. Diddams, "Microresonator-based optical frequency combs." Science 332, 555-559 (2011).

[13] F. Bo, J. Wang, J. Cui, Ş. K. Özdemir, Y. Kong, G. Zhang, J. Xu, and L. Yang, "Lithium-Niobate-Silica hybrid whispering-gallery-mode resonators." Adv. Mater. 27, 8075--8081 (2015).

[14] L. Yang, D. K. Armani, and K. J. Vahala, "Fiber-coupled erbium microlasers on a chip." Appl. Phys. Lett. 83, 825 (2003).

[15] T. J. Kippenberg, J. Kalkman, A. Polman, and K. J. Vahala, "Demonstration of an erbium-doped microdisk laser on a silicon chip." Phys. Rev. A 74, 051802(R) (2006).

[16] L. He, Ş. K. Özdemir, J. Zhu, W. Kim, and L. Yang, "Detecting single viruses and nanoparticles using whispering gallery microlasers." Nat. Nanotechnol. 6, 428 (2011).

[17] Ş. K. Özdemir, J. Zhu, X. Yang, B. Peng, H. Yilmaz, L. He, F. Monifi, S. H. Huang, G. Long, and L. Yang, "Highly sensitive detection of nanoparticles with a self-referenced and self-heterodyned whispering-gallery raman microlaser." Proc. Natl. Acad. Sci. USA 111(37), E3836-E3844 (2014).

[18] B.-B. Li, W. R. Clements, X.-C. Yu, K. Shi, Q. Gong, and Y.-F. Xiao, "Single nanoparticle detection using split-mode microcavity raman Lasers." Proc. Natl. Acad. Sci. USA 111(41), 14657-14662 (2014).

[19] Z.-P. Liu, J. Zhang, Ş. K. Özdemir, B. Peng, H. Jing, X.-Y., Lü, C.-W. Li, L. Yang, F. Nori, and Y.-X. Liu, "Metrology with PT-Symmetric Cavities: Enhanced Sensitivity near the PT-Phase Transition." Phys. Rev. Lett. 117, 110802 (2016).

[20] M. Asano, K. Y. Bliokh, Y. P. Bliokh, A. G. Kofman, R. Ikuta, T. Yamamoto, Y. S. Kivshar, L. Yang, N. Imoto, Ş. K. Özdemir, and F. Nori "Anomalous time delays and quantum weak measurements in optical micro-resonators." Nat. Commun. 7, 13488 (2016).

[21] B. Peng, Ş. K. Özdemir, F. Lei, F. Monifi, M. Gianfreda, G.-L. Long, S. Fan, F. Nori, C. M. Bender, and L. Yang, "Parity-time-symmetric whispering-gallery microcavities." Nature Phys. 10, 394--398 (2014).

[22] L. Chang, X. Jiang, S. Hua, C. Yang, J. Wen, L. Jiang, G. Li, G. Wang, and M. Xiao, "Parity-time symmetry and variable optical isolation in active-passive-coupled microresonators." Nat. Photonics 5, 524--529 (2014).

[23] B. Peng, Ş. K. Özdemir, S. Rotter, H. Yilmaz, M. Liertzer, F. Monifi, C. M. Bender, F. Nori, and L. Yang, L. "Loss-induced suppression and revival of lasing." Science 346, 328-332 (2014).

[24] H. Jing, Ş. K. Özdemir, X.-Y. Lü, J. Zhang, L. Yang, and F. Nori, "PT-Symmetric Phonon Laser." Phys. Rev. Lett. 113, 053604 (2014).

[25] H. Hodaei, M.-A. Miri, M. Heinrich, D. N. Christodoulides, and M. Khajavikhan, "Parity-time-symmetric microring lasers." Science 346, 975-978 (2014).

[26] S. Weis, R. Riviere, S. Deleglise, E. Gavartin, O. Arcizet, A. Schliesser, and T. J. Kippenberg, "Optomechanically Induced Transparency." Science 330, 1520-1523 (2010).

[27] Z. Shen, Y.-L. Zhang, Y. Chen, C.-L. Zou, Y. -F. Xiao, X.-B. Zou, F.-W. Sun, G.-C. Guo, and C.-H. Dong, "Experimental realization of optomechanically induced non-reciprocity." Nature Photon. 10, 657-661 (2016).

[28] B. Peng, Ş. K. Özdemir, W. Chen, F. Nori, and L. Yang, "What is and what is not electromagnetically-induced-transparency in whispering-gallery-microcavities." Nat. Commun. 5, 5082 (2014).

[29] C. Dong, V. Fiore, M. C. Kuzyk, and H. Wang, "Optomechanical dark mode." Science 338, 1609--1613 (2012).

[30] L. Fan, K. Y. Fong, M. Poot, and H. X. Tang, "Cascaded optical transparency in multimode-cavity optomechanical systems." Nat. Commun. 6, 5850 (2015).

[31] X. Yang, Ş. K. Özdemir, B. Peng, H. Yilmaz, F.-C. Lei, G.-L. Long, and L. Yang, "Raman gain induced mode evolution and on-demand coupling control in whispering-gallery-mode microcavities." Opt. Express 23, 29573--29583 (2015).

[32] K. Totsuka, and M. Tomita, "Optical microsphere amplification system." Opt. Lett. 32, 3197--3199 (2007).

[33] V. Huet, A. Rasoloniaina, P. Guilleme, P. Rochard, P. Feron, M. Mortier, A. Levenson, K. Bencheikh, A. Yacomotti, and Y. Dumeige, "Millisecond Photon Lifetime in a Slow-Light Microcavity." Phys. Rev. Lett. 116, 133902 (2016).

[34] A. Rasoloniaina, V. Huet, T. K. N. Nguyen, E. L. Cren, M. Mortier, L. Michely, Y. Dumeige, and P. Feron, "Controling the coupling properties of active ultrahigh-Q WGM microcavities from undercoupling to selective amplification." Sci. Rep. 4, 4023 (2014).

[35] X.-F. Liu, F. Lei, M. Gao, X. Yang, C. Wang, Ş. K. Özdemir, L. Yang, and G.-L. Long, "Gain competition induced mode evolution and resonance control in erbium-doped whispering-gallery microresonators." Opt. Express 24, 9550--9560 (2016).

[36] L. Yang, and K. J. Vahala, "Gain functionalization of silica microresonators." Opt. Lett. 28, 592-594 (2003).

[37] S. M. Spillane, T. J. Kippenberg, O. J. Painter, and K. J. Vahala, "Ideality in a Fiber-Taper-Coupled Microresonator System for Application to Cavity Quantum Electrodynamics." Phys. Rev. Lett. 91, 043902 (2003).

[38] Ming Cai, Oskar Painter, and Kerry J. Vahala, "Observation of Critical Coupling in a Fiber Taper to a Silica-Microsphere Whispering-Gallery Mode System." Phys. Rev. Lett. 85, 74 (2000).

[39]T. Carmon, L. Yang, and K. J. Vahala, "Dynamical thermal behavior and thermal self-stability of microcavities." Opt. Express 12, 4742--4750 (2004).

[40] V. B. Braginsky, M. L. Gorodetsky, and V. S. Ilchenko, "Quality-factor and nonlinear properties of optical whispering-gallery modes." Phys. Lett. A 137, 393-397 (1989).

[41] E. M. Purcell, "Spontaneous emission probabilities at radio Frequencies." Phys. Rev. 69, 681 (1946).